\documentclass[10pt]{article}

\usepackage[a4paper, margin=1in]{geometry}

\usepackage{amsmath, amssymb}

\usepackage{url}
\usepackage{graphicx}
\usepackage{xcolor}
\usepackage[normalem]{ulem}

\usepackage{authblk}

\usepackage[hidelinks]{hyperref}

\title{Pendulum Tracker -- SimuFísica\textsuperscript{\textregistered}: A Web-based Tool for Real-time Measurement of Oscillatory Motion}

\author[1]{Marco P. M. de Souza}
\author[1]{Juciane G. Maia}
\author[2]{Lilian N. de Andrade}

\affil[1]{Departamento de Física, Universidade Federal de Rondônia, 76900-726, Ji-Paraná, Rondônia, Brazil}
\affil[2]{Secretaria de Estado da Educação de Rondônia, 78903-043, Porto Velho, Rondônia, Brazil}

\date{}

\begin{document}
	
\maketitle

\begin{abstract}
We present Pendulum Tracker, a computer vision-based application that enables real-time measurement of the oscillatory motion of a physical pendulum. Integrated into the educational platform SimuFísica, the system uses the OpenCV.js library and runs directly in the browser, working on computers, tablets, and smartphones. The application automatically detects the pendulum's position via the device's camera, displaying in real time the angle-versus-time graph and estimates of the oscillation period. Experimental case studies demonstrate its effectiveness in measuring the period, determining gravitational acceleration, and analyzing damped oscillations. The results show excellent agreement with theoretical predictions, confirming the system's accuracy and its applicability in educational contexts. The accessible interface and the ability to export raw data make Pendulum Tracker a versatile tool for experimental physics teaching.
\end{abstract}

\vspace{2pc}
\noindent{\it Keywords}: computer vision, pendulum motion, physics education, real-time measurement, web-based application

\section{Introduction}

The study of the pendulum plays a central role in physics education, offering an accessible and visually clear way to introduce fundamental concepts such as harmonic motion, energy conservation, damping, and the determination of gravitational acceleration. Due to its conceptual simplicity and ease of experimental setup, the pendulum has been widely used in classrooms and educational laboratories at different educational levels \cite{Oliveira2016, Dandare2018, Lacsny2014}. Its use also extends to more advanced investigations, including the Foucault pendulum \cite{Salva}, nonlinear oscillations, chaotic dynamics, and dissipative effects \cite{Hinrichsen2020}.

In recent decades, digital tools such as the Tracker software have been employed for experimental physics studies involving video analysis \cite{Pedersen2020, Cross-2025}. Tracker enables data collection from camera-captured images, promoting greater interactivity and conceptual understanding. However, its use typically requires prior recording and often manual frame-by-frame analysis, which can limit its adoption in some educational settings --- especially for real-time experiments.

In this work, we present Pendulum Tracker, a computer vision-based application integrated into the SimuFísica educational platform. Designed to run directly in the browser, the system allows automatic, real-time measurement of the angular motion of simple pendulums. Section~\ref{pendulum-tracker} provides an overview of the Pendulum Tracker and its features. Section~\ref{examples} presents three usage examples: oscillation period measurement, experimental determination of gravitational acceleration, and analysis of damped oscillations. The results demonstrate the tool's robustness and its usefulness as a complementary resource for experimental physics teaching at both high school and undergraduate levels.

\section{The Pendulum Tracker -- SimuFísica\textsuperscript{\textregistered}}
\label{pendulum-tracker}

\subsection{Overview}

Pendulum Tracker is a tool developed for the detailed experimental study of the oscillatory motion of pendulums, using computer vision to capture and analyze the angular position of the object in real time. Based on the OpenCV.js library \cite{opencv}, an implementation of OpenCV for JavaScript, Pendulum Tracker is integrated into the SimuFísica platform, which offers a wide range of applications and simulators aimed at teaching and learning physics in areas such as mechanics \cite{Souza-2024-2-x}, electromagnetism \cite{Souza-2024-1-x}, thermodynamics \cite{Souza-2025-2}, and quantum mechanics \cite{Souza-2025-1}.

Designed as a web-based application, Pendulum Tracker is easily accessible through any modern browser and is compatible with various devices such as computers, tablets, and smartphones, requiring only an integrated or external camera for video capture. Although the Pendulum Tracker is already available in the online version of SimuFísica, it is not yet included in the offline versions distributed through app stores. A future update will add the application to the downloadable packages for Android (Play Store), iOS (App Store), Windows (Microsoft Store), and Linux (Snapcraft), enabling its use without an internet connection.

\subsection{Interface and functionality}

\begin{figure}[ht]
	\centering
	\includegraphics[width=0.50\linewidth]{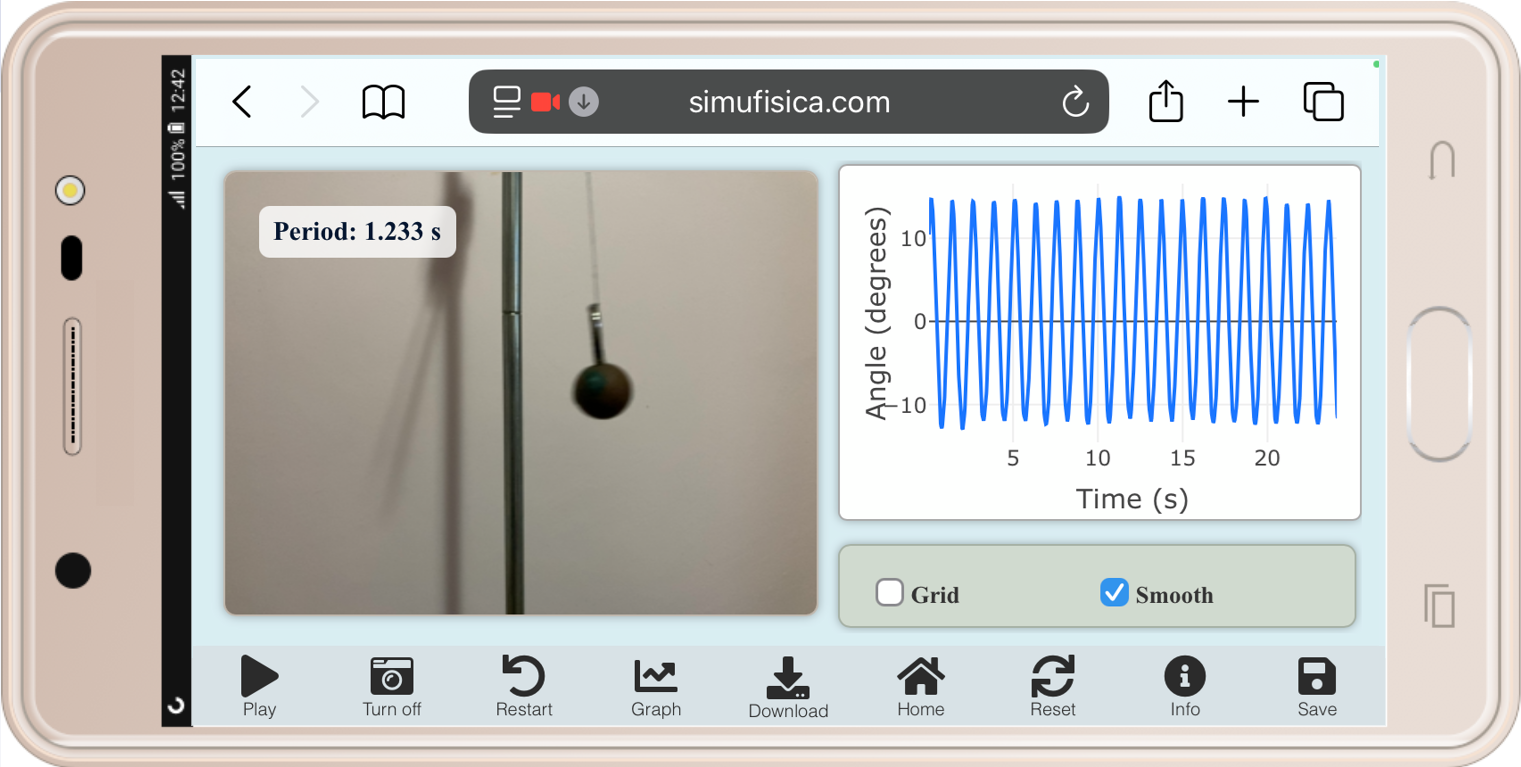}
	\caption{Pendulum Tracker application from the SimuFísica\textsuperscript{\textregistered} platform. Access link: \url{https://simufisica.com/en/pendulum-tracker/}.}
	\label{fig1}
\end{figure}

Figure~\ref{fig1} shows the application interface in operation. After fixing the capture device on a stable support and properly aiming it at the pendulum, the user activates the camera by clicking the \texttt{Turn on} button located on the toolbar. Then, the user must set the pendulum into motion, and it is recommended to maintain small angles (less than 20\textdegree) to ensure agreement with the theoretical expression for the period $T$ of a simple pendulum of length $L$:

\begin{equation}
	\label{eq-T}
	T = 2\pi \sqrt{\frac{L}{g}},
\end{equation}

\noindent where $g$ is the local acceleration due to gravity. It is also recommended to use a thin black string to minimize interference with automatic detection, and to center the pendulum's motion in the frame, occupying a good portion of the screen. To assist with this positioning, the user can enable the \texttt{Grid} checkbox, which displays a reference mesh on the screen. The background does not need to be uniform, and the pendulum object does not need to be spherical --- tests conducted indicate excellent results in domestic environments using a pendulum composed of a cylindrical object or even a small padlock. One requirement is that no other motion occurs in the background besides the pendulum's oscillation.

After the initial setup, the user clicks the \texttt{Start} button. A short calibration interval (approximately 3 s) begins, during which the maximum and minimum positions of the pendulum are collected so that an angular estimate can be provided on the vertical axis of the graph. After this interval, the angle versus time curve is displayed on the screen. Shortly afterward, the user can observe the pendulum's period, obtained from the average of five oscillation cycles. Finally, for more precise measurements, as shown in Section~\ref{examples}, the user has the option to download the curve data using the \texttt{Download} button for analysis in spreadsheets such as OriginLab, Microsoft Excel, or similar software.

\section{Application examples}
\label{examples}

In this section, we present three representative examples that illustrate the capabilities of the Pendulum Tracker application from the SimuFísica platform. These activities can be used in the classroom as practical tasks or integrated into home-based experimental work.

\subsection{Period measurement}
\label{period-measurement}

\begin{figure}[ht]
	\centering
	\includegraphics[width=0.50\linewidth]{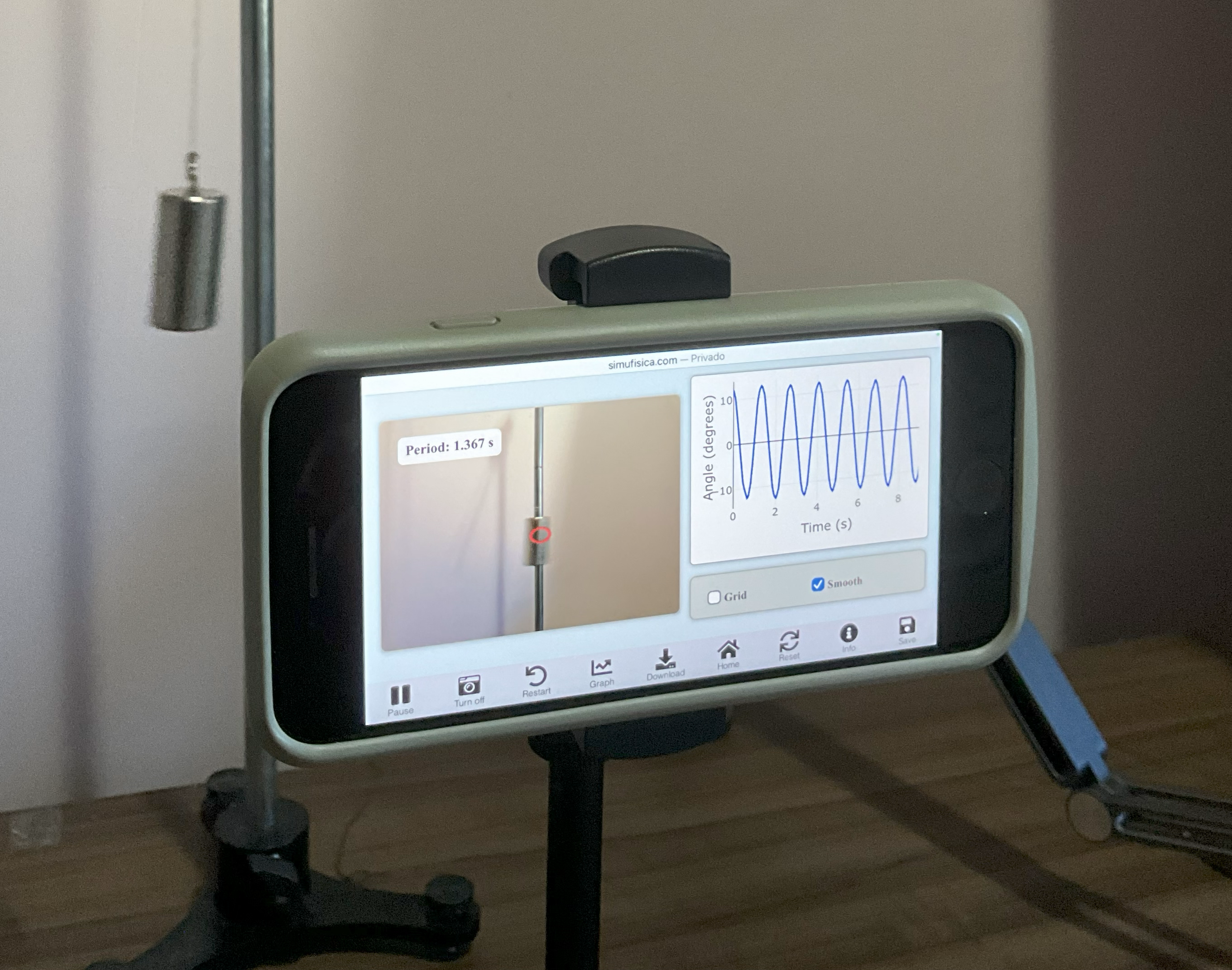}
	\caption{Measurements performed using the Pendulum Tracker -- SimuFísica application with $L = 45.8 \pm 0.1$ cm.}
	\label{fig2}
\end{figure}

In the first example, we used a smartphone (iPhone SE, 2020 model) mounted on a generic holder (Figure~\ref{fig2}). As the pendulum mass, we used an aluminum cylinder with a hook. The pendulum length was measured with a tape measure and determined to be $L = 45.8 \pm 0.1$ cm, from the point of support to the cylinder's center of mass. After a few oscillation cycles, the application began displaying the average period on the screen, with observed fluctuations between $T = 1.367$ s and $T = 1.374$ s. We compared these results with the theoretical value given by Eq.~(\ref{eq-T}). The measurements were performed in the city of Ji-Paraná, in the state of Rondônia, Brazil, located at latitude $\phi\times 180/\pi = -10.88$\textdegree. Considering $g = 9.7821671$ m/s$^2$ as the local gravitational acceleration, obtained from the equation \cite{Hinze}

\begin{equation}
	g(\phi) = 9.780327\left[ 1 + 0.0053024\sin^2(\phi) - 0.0000058\sin^2(2\phi) \right] \mbox{m/s}^2,
	\label{eq-g}
\end{equation}

\noindent we found a maximum error of only 1.1\%.

\begin{figure}[ht]
	\centering
	\includegraphics[width=0.95\linewidth]{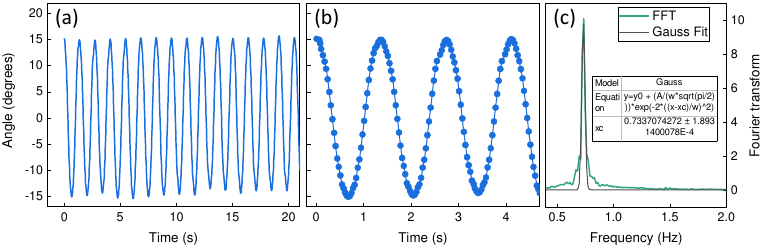}
	\caption{Results from the measurements shown in Fig. \ref{fig2}. (a) Angle as a function of time. (b) Zoom of (a) within a 4.7 s window. The line is a guide to the eye. (c) Green line shows the Fourier transform of (a), considering a 60 s interval. The black line is a Gaussian fit of the Fourier transform.}
	\label{fig3}
\end{figure}

Figure~\ref{fig3} presents a more detailed analysis of the obtained data. In panel (a), we show the time evolution of the pendulum angle. Panel (b) shows a zoom of the first 4.7 s, revealing a sinusoidal curve with approximately 40 data points per cycle --- this number depends on the device’s frame rate and the pendulum length. The application's algorithm automatically smooths the data to reduce noise, but the user can disable this by unchecking the \texttt{Smooth} checkbox.

Figure~\ref{fig3}(c) shows, in green, the spectrum obtained from the Fourier transform of the data in panel (a), using a 60-second time window. A Gaussian fit to the main peak yields a frequency of $f = 0.7337 \pm 0.0002$ Hz, corresponding to an experimental period of

\begin{equation}
	T_{\mathrm{exp}} = 1.3630 \pm 0.0004 \mbox{ s},
\end{equation}

\noindent which is very close to the theoretical value,

\begin{equation}
	T_{\mathrm{theo}} = 1.3595 \mbox{ s},
\end{equation}

\noindent although slightly outside the uncertainty margin.

\subsection{Measurement of the acceleration due to gravity}

In this second example, we used Pendulum Tracker to determine the local gravitational acceleration based on the dependence between the period $T$ and the pendulum length $L$. To that end, we conducted measurements using different values of $L$. The procedure was similar to that in Section~\ref{period-measurement}: after each measurement, the data were exported, a Fourier transform was applied, and a Gaussian fit was performed to obtain the central frequency and calculate the period.

\begin{figure}[ht]
	\centering
	\includegraphics[width=0.85\linewidth]{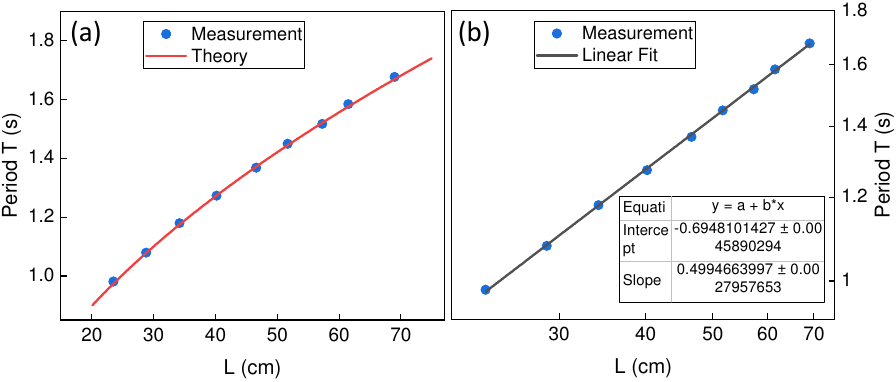}
	\caption{Pendulum oscillation period as a function of length. (a) Linear scale: experimental data in blue and theoretical curve in red. (b) Log-log plot of the experimental data (blue) with linear fit (black line).}
	\label{fig4}
\end{figure}

The results are presented in Figure~\ref{fig4}. Panel (a), in linear scale, shows good agreement between the experimental data (blue) and the theoretical curve (red). Panel (b), in log-log scale, reveals a power-law relationship $T = C L^n$, with a linear fit yielding

\begin{equation}
	n_{\mathrm{exp}} = 0.499 \pm 0.003,
\end{equation}

\noindent which is consistent with the theoretical value $n_{\mathrm{theo}} = 0.5$.

From the linear fit intercept $a = -0.695 \pm 0.005$ in Fig.~\ref{fig4}(b), and the relation $T = C L^n$, we obtain

\begin{equation}
	g = \left( \frac{2\pi}{10^{a+1}} \right)^2,
\end{equation}

\noindent with uncertainty given by

\begin{equation}
	\Delta g = \frac{8\pi^2 \ln(10)}{10^{2(a+1)}} \Delta a,
\end{equation}

\noindent where $\Delta a$ is the uncertainty in the value of $a$. Using the data from Fig.~\ref{fig4}(b), we find

\begin{equation}
	g_{\mathrm{exp}} = 9.7 \pm 0.2 \mbox{ m/s}^2,
\end{equation}

\noindent which is in excellent agreement with the expected local value ($g_{\mathrm{theo}} = 9.78$ m/s$^2$).

\subsection{Damped oscillations --- experimente vs. theory}

\begin{figure}[ht]
	\centering
	\includegraphics[width=0.99\linewidth]{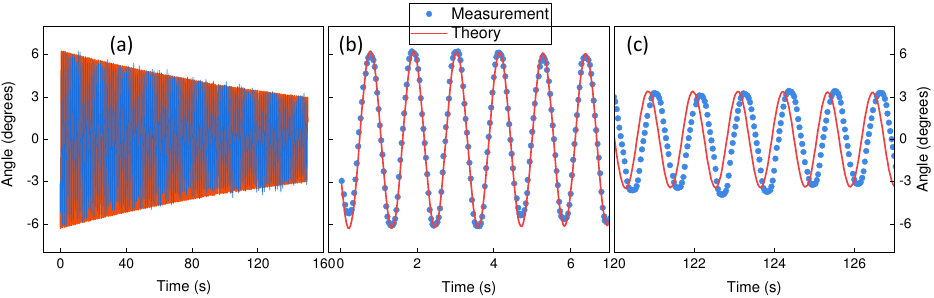}
	\caption{(a) Damped oscillations of a simple pendulum. Experimental data in blue and theoretical curve in red. (b) Zoom of (a) up to $t = 7$ s. (c) Zoom of (a) in the time window $t \in [120, 127]$ s.}
	\label{fig5}
\end{figure}

In the final example, we explored Pendulum Tracker's ability to record damped oscillations over extended time intervals. We used a pendulum of length $L = 31.2 \pm 0.1$ cm, collecting data from $t = 0$ to $t = 150$ s, resulting in 4366 data points (Fig.~\ref{fig5}). This data density demonstrates one of the main advantages of Pendulum Tracker over traditional tools like Tracker \cite{tracker}, which often require manual frame-by-frame analysis. Moreover, the SimuFísica platform's application provides the oscillation period and angle graph in real time on widely used devices such as smartphones.

The comparison with the theoretical model is based on the harmonic approximation for the motion of a damped pendulum \cite{Quiroga}:

\begin{equation}
	\theta(t) = \theta_0e^{-\gamma t}\cos\left( t\sqrt{\frac{g}{L} - \gamma^2} - \delta\right),
	\label{theory}
\end{equation}

\noindent where $g = 9.782167$ m/s$^2$, $\theta_0 = 6.3^\circ$, $\gamma = 0.005$ s$^{-1}$, and $\delta = 1.4\pi$. The values of $\theta_0$, $\gamma$, and $\delta$ were chosen to maximize the agreement between the theoretical curve and the experimental data in the interval $t \in [0, 7]$ s [Fig.~\ref{fig5}(b)]. For longer times [Fig.~\ref{fig5}(c)], we observe a slight phase shift between the model and the data, likely due to small variations in ambient conditions.

\section{Conclusions}

In this work, we presented Pendulum Tracker, a computer vision-based application that allows precise real-time measurements of a pendulum's oscillatory motion using an ordinary device camera. The tool proved effective both for verifying well-known experimental relations, such as the period equation and its dependence on gravitational acceleration, and for more advanced investigations, such as the study of damped oscillations. The good agreement with the theory taught at high school and undergraduate levels confirms the application's reliability.

As future work, we plan to expand the application's functionalities, including support for different experimental configurations, such as double\footnote{\url{https://simufisica.com/en/double-pendulum/}} or coupled pendulums\footnote{\url{https://simufisica.com/en/coupled-pendulums/}}. In addition, we are updating the SimuFísica app store versions to include the Pendulum Tracker tool, thereby enabling offline use in educational environments with limited internet access.

\section*{Data availability statement}

All data from this work are available in the repository at \url{https://zenodo.org/records/15569631}.

\section*{Acknowledgements}

This research was funded by Fundação de Amparo ao Desenvolvimento das Ações Científicas e Tecnológicas e à Pesquisa do Estado de Rondônia (FAPERO, Grant 36214.577.20546.20102023) and Universidade Federal de Rondônia (UNIR, Grant 23118.006316/2024-79). M. P. M. de Souza acknowledges financial support from Conselho Nacional de Desenvolvimento Científico e Tecnológico (CNPq, Grant 304017/2022-1).

\bibliographystyle{iopart-num}
\bibliography{biblio}

\providecommand{\noopsort}[1]{}\providecommand{\singleletter}[1]{#1}%
\providecommand{\newblock}{}
\begin{thebibliography}{10}
\expandafter\ifx\csname url\endcsname\relax
  \def\url#1{{\tt #1}}\fi
\expandafter\ifx\csname urlprefix\endcsname\relax\def\urlprefix{URL }\fi
\providecommand{\href}[2]{#1}  
\providecommand{\eprint}[2][arXiv]{#1:\linebreak[0]#2}

\bibitem{Oliveira2016}
Oliveira V 2016 {\em Physics Education\/} {\bf 51} 063007

\bibitem{Dandare2018}
Dandare K 2018 {\em Physics Education\/} {\bf 53} 055002
  \urlprefix\url{https://dx.doi.org/10.1088/1361-6552/aac92f}

\bibitem{Lacsny2014}
Lacsny B, {\v{S}}tubna I and Teleki A 2014 {\em European Journal of Physics\/}
  {\bf 35} 065023

\bibitem{Salva}
Salva N~N and Salva H~R 2025 {\em American Journal of Physics\/} {\bf 93}
  297--307 ISSN 0002-9505 \urlprefix\url{https://doi.org/10.1119/5.0208092}

\bibitem{Hinrichsen2020}
Hinrichsen P~F 2020 {\em European Journal of Physics\/} {\bf 41} 055002

\bibitem{Pedersen2020}
Pedersen H~B, Andersen J~E~V, Nielsen T~G, Iversen J~J, Lyckegaard F and
  Mikkelsen F~K 2020 {\em European Journal of Physics\/} {\bf 41} 015701

\bibitem{Cross-2025}
Cross R 2025 {\em Physics Education\/} {\bf 60} 033005
  \urlprefix\url{https://dx.doi.org/10.1088/1361-6552/adbed0}

\bibitem{opencv}
 2000 {OpenCV}: Open source computer vision library \url{https://opencv.org/}

\bibitem{Souza-2024-2-x}
Souza M~P~M, Oliveira C~M and Araújo R~P~P 2024 {\em arXiv:2410.17951
  [physics.ed-ph]\/} This is the English version of a peer-reviewed paper
  published in Portuguese: A Física na Escola 22, 240173 (2024). DOI:
  \url{https://doi.org/10.59727/fne.v22i1.173}
  \urlprefix\url{https://doi.org/10.48550/arXiv.2410.17951}

\bibitem{Souza-2024-1-x}
Souza M~P~M, Oliveira S~P and Luiz V 2024 {\em arXiv:2410.18721
  [physics.ed-ph]\/} This is the English version of a peer-reviewed paper
  published in Portuguese: Rev. Bras. Ensino Fís. 46, e20230219 (2024). DOI:
  \url{https://doi.org/10.1590/1806-9126-RBEF-2023-0219}
  \urlprefix\url{https://doi.org/10.48550/arXiv.2410.18721}

\bibitem{Souza-2025-2}
Souza M~P~M and Siqueira A~B 2025 {\em arXiv:2506.03289 [physics.ed-ph]\/}
  \urlprefix\url{https://doi.org/10.48550/arXiv.2506.03289}

\bibitem{Souza-2025-1}
Souza M~P~M, Pavão G~H~H, de~Almeida A~A~C and Vianna S~S 2025 {\em
  arXiv:2506.01108 [quant-ph]\/}
  \urlprefix\url{https://doi.org/10.48550/arXiv.2506.01108}

\bibitem{Hinze}
Hinze W~J, von Frese R~R~B and Saad A~H 2013 {\em Gravity and Magnetic
  Exploration: Principles, Practices, and Applications\/} (Cambridge University
  Press)

\bibitem{tracker}
Brown D, Christian W and Hanson R~M 2025 Tracker: Video analysis and modeling
  tool \url{https://opensourcephysics.github.io/tracker-website/}

\bibitem{Quiroga}
Quiroga G~D and Ospina-Henao P~A 2017 {\em European Journal of Physics\/} {\bf
  38} 065005 \urlprefix\url{https://dx.doi.org/10.1088/1361-6404/aa8961}

\end{thebibliography}

\end{document}